\documentclass[letter]{aa} 

\usepackage{graphicx}   
\usepackage{amsmath}    
\usepackage{amsmath, amsfonts, amssymb, graphics,graphicx, wrapfig, nicefrac}
\usepackage{epsfig}
\usepackage{epstopdf}
\usepackage{multirow}
\usepackage{multicol}
\usepackage{graphicx}
\usepackage{rotating}
\usepackage{mathrsfs}
\graphicspath{{./}{Figures/}}
\usepackage{txfonts}

\usepackage{soul}

\usepackage{color}

\begin{document}

   \title{
    The CO-dark molecular gas in the cold H\,{\sc i} arc}

   \author{Gan Luo
          \inst{1}
          \and
          Di Li\inst{2,3}
          \and
          Zhi-Yu Zhang\inst{4,5}
          \and 
          Thomas G. Bisbas\inst{6}
          \and
          Ningyu Tang\inst{7}
          \and
          Lingrui Lin\inst{4,5}
          \and
          Yichen Sun\inst{4,5}
          \and
          Pei Zuo\inst{2}
          \and
          Jing Zhou\inst{4,5}
          }

   \institute{Institut de Radioastronomie Millimetrique, 300 rue de la Piscine, 38400, Saint-Martin d’Hères, France\\
              \email{luo@iram.fr}
        \and
             CAS Key Laboratory of FAST, National Astronomical Observatories, Chinese Academy of Sciences, Beijing 100101, China
        \and
             University of Chinese Academy of Sciences, Beijing 100049, China
         \and
             School of Astronomy and Space Science, Nanjing University, Nanjing 210093, People’s Republic of China
        \and
            Key Laboratory of Modern Astronomy and Astrophysics (Nanjing University), Ministry of Education, Nanjing 210093, People’s Republic of China
        \and
            Research Center for Astronomical Computing, Zhejiang Laboratory, Hangzhou 311100, China
        \and
            Department of Physics, Anhui Normal University, Wuhu, Anhui 241002, China
             }

   \date{Received xx; accepted xx}

\abstract{The CO-dark molecular gas (DMG), which refers to the molecular gas not traced by CO emission, is crucial for the evolution of the interstellar medium (ISM). While the gas properties of DMG have been widely explored in the Solar neighborhood, whether or not they are similar in the outer disk regions of the Milky Way is still not well understood. In this Letter, we confirm the existence of DMG toward a cold H\,{\sc i} arc structure at 13 kpc away from the Galactic center with both OH emission and H\,{\sc i} narrow self-absorption (HINSA). This is the first detection of HINSA in the outer disk region, in which the HINSA fraction ($N_{\rm HINSA}$/$N_{\rm H_2}$ = 0.022$\pm$0.011) is an order of magnitude higher than the average value observed in nearby evolved dark clouds, but is consistent with that of the early evolutionary stage of dark clouds. The inferred H$_2$ column density from both extinction and OH emission ($N_{\rm H_2} \approx 10^{20}$\,cm$^{-2}$) is an order of magnitude higher than previously estimated. 
Although the ISM environmental parameters are expected to be different between the outer Galactic disk regions and the Solar neighborhood, we find that the visual extinction ($A_{\rm V}$ = 0.19$\pm$0.03\,mag), H$_2$-gas density ($n_{\rm H_2} = 91\pm46$\,cm$^{-3}$), and molecular fraction (58\%$\pm$28\%) of the DMG are rather similar to those of nearby diffuse molecular clouds.
The existence of DMG associated with the expanding H\,{\sc i} supershell supports a scenario where the expansion of supershells may trigger the formation of molecular clouds within a crossing timescale of the shock wave ($\sim$10$^6$\,yr).
}



   \maketitle
%

\section{Introduction} \label{sec:intro}

The recycling between interstellar matter and stars is important for understanding star formation and galaxy evolution. H\,{\sc i} and H$_2$ are the two most important constituents of the interstellar medium (ISM).
While H\,{\sc i} can be detected directly by the 21\,cm emission line \citep{McKee1977,Dickey1990,Heiles2003,Kalberla2009}, H$_2$ cannot be detected directly in emission in the cold environment (e.g., 15\,K of molecular clouds) because of the large energy level spacing and weak transition strengths. CO has been used extensively to probe the morphology and physical properties of cold molecular gas over the past half-century \citep{Wilson1970,Dame2001,Bolatto2013}.

Growing evidence from $\gamma$-ray \citep[e.g., EGRET,
$Fermi$,][]{Grenier2005,Remy2017} and dust thermal emission \citep[e.g., \textit{Planck},][]{Planck2011} over the past 20 years shows that nearly half of the total neutral gas cannot be traced by H\,{\sc i} and CO emission. Studies focusing on spectral lines (e.g., spectra of C$^+$, OH, HCO$^+$) further strengthen the hypothesis that the ``unseen'' gas is mostly CO-dark molecular gas \citep[DMG,][]{Pineda2013,Lee2015,Li2018,Luo2020,Busch2021,Rybarczyk2022}.

DMG is most likely to exist in the low-to-intermediate extinction range ($A_{\rm V}$: 0.2--2\,mag), where the cloud is going through a H\,{\sc i}-to-H$_2$ transition phase \citep{Seifried2020}. The fraction of DMG increases monotonically with Galactocentric distance \citep[$R_{\rm GC}$,][]{Pineda2013}, which is consistent with simulations that show that the fraction of DMG decreases with metallicity \citep{Hu2021}. However, the threshold (e.g., $A_{\rm V}$, $n_{\rm H}$) at which H\,{\sc i} transforms into H$_2$ in the outer disk region is still unclear.

An extremely cold ($T_{\rm kin}$ down to 10\,K), extended (spatial scale $L\sim2\,{\rm kpc}$), and massive ($M\sim$1.9$\times$10$^7$M$_{\odot}$) H\,{\sc i} arc structure (red curves in Figure \ref{fig:arc}) was found through H\,{\sc i} self-absorption (HISA) signatures without associated CO emission \citep{Knee2001}. This structure is located in the northern rim of the expanding supershell GSH139-03-69 \citep[$R_{\rm GC}$ $\sim$16\,kpc\footnote{We revise the distance to 13\,kpc according to the latest rotational curve by \citet{Reid2019}.},][]{Heiles1979}, but is not supposed to exist according to the classical ISM phase models \citep{Field1969,McKee1977}. As the cooling of H\,{\sc i} at such low temperatures is extremely difficult, the H\,{\sc i} arc structure could either be in an unstable adiabatic expansion \citep{Sahai1997,Sahai2013}, or be coexistent with unseen DMG.

In this Letter, we present high-sensitivity and high-spectral-resolution H\,{\sc i} observations with the Five-hundred-meter Aperture Spherical Telescope (FAST), and OH observations with the {\it Robert C. Byrd} Green Bank Telescope (GBT) toward the arc structure. The H\,{\sc i} narrow self-absorption (HINSA) feature and the OH emission confirm the existence of DMG within the H\,{\sc i} arc. We highlight the similar physical conditions and H\,{\sc i}-to-H$_2$ transition for both nearby clouds and outer disk regions, and the potential of HINSA to probe the DMG properties in future works. 

\section{Observations}\label{sec:obs}

\subsection{FAST H\,{\sc i} observations}\label{sec:hi data}

\begin{figure}
\includegraphics[width=1.0\linewidth]{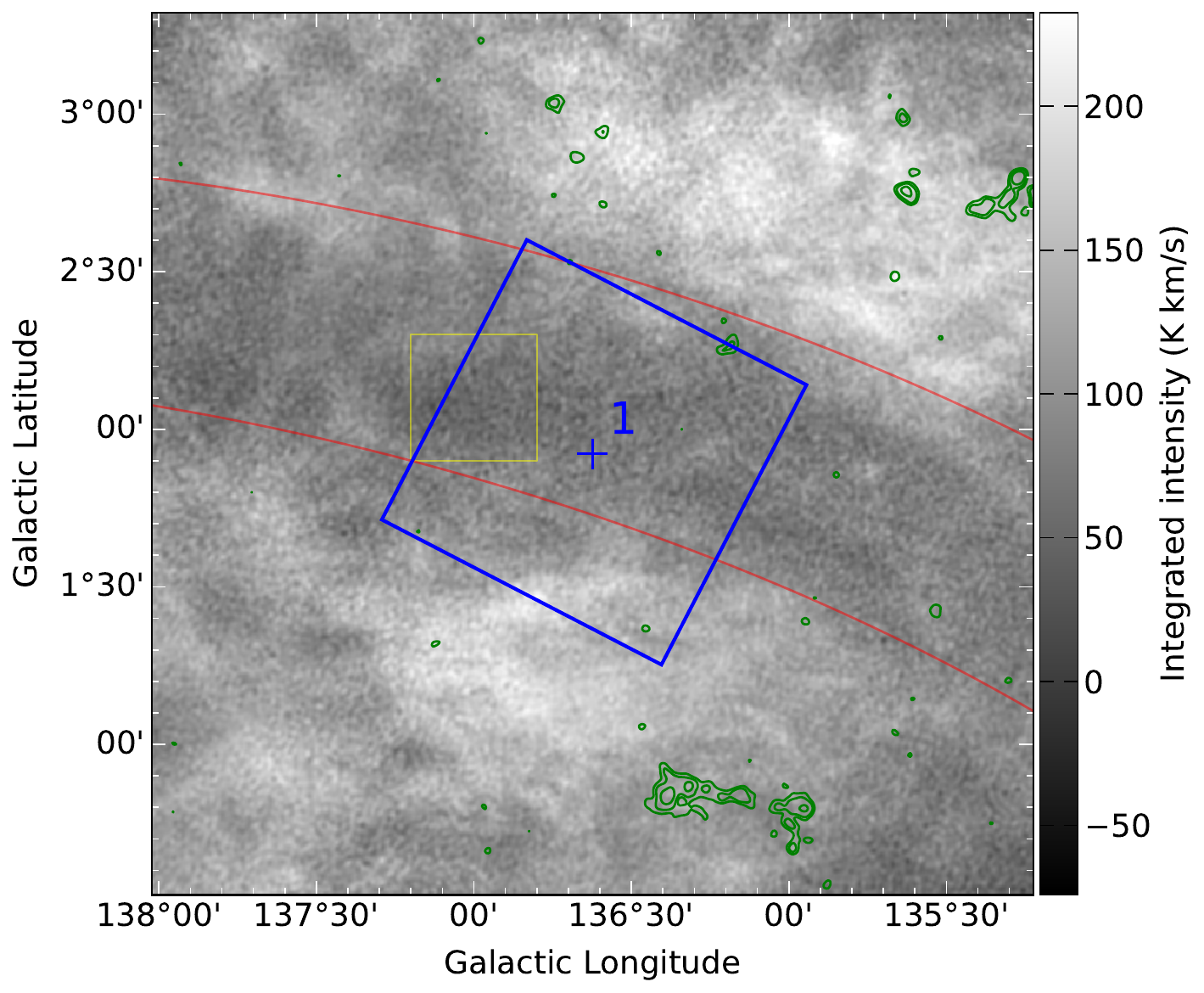}
\caption{H\,{\sc i} integrated intensity (background gray scale) from the Canadian Galactic Plane Survey \citep[CGPS,][]{Taylor2003} and CO integrated intensity (green contours, 0.4 K\,km\,s$^{-1}$ $\times$ (3, 5, 10)) from the Outer Galaxy Survey \citep[OGS,][]{Heyer1998}. The integrated velocity range is $-$80 $\sim$ $-$73\,km\,s$^{-1}$ for both H\,{\sc i} and CO. The H\,{\sc i} arc is denoted by the red curves. The yellow square denotes the position where the inferred H\,{\sc i} spin temperature is $\sim$10\,K in \citet{Knee2001}. The blue cross denotes the H\,{\sc i} pointing observations, and the blue square is the region mapped in OH.} 
\label{fig:arc}
\end{figure}

Single-pointing H\,{\sc i} observations (blue cross in Fig.~\ref{fig:arc}) were made using the central beam of the 19-beam receiver of FAST ---with the total power ON mode--- in December 2019 (PI: Ningyu Tang). The rest frequency for the H\,{\sc i} observation is 1420.2058\,MHz and the spectral resolution is 473\,Hz, corresponding to a velocity resolution of 0.1\,km\,s$^{-1}$ at 1.4\,GHz. The system temperature $T_{\rm sys}$ is 22 K and the integration time is 600\,s. The root mean square (rms) of the H\,{\sc i} spectrum (in $T\rm_A$ units) is 0.054\,K per 0.1\,km\,s$^{-1}$. The main beam efficiency of $\eta = 0.87$ was adopted to convert $T\rm_A$ into brightness temperature $T\rm_{mb}$ for the FAST central beam \citep{2020RAA....20...77T}.

\subsection{GBT OH observations}\label{sec:hcop data}

We performed mapping observations of four OH $\Lambda$-doubling lines in the $^2\Pi_{3/2}$, $J$ = 3/2 energy level from Feb 10, 2022 to March 12, 2022, using the L-band receiver on board the GBT in on-the-fly mode (Project ID: 22A-228, PI: Gan Luo). The observed map is centered at right ascension (RA) = 02$^h$50$^m$40.1$^s$ and declination (Dec) = +61$^\circ$34$'$58.3$''$, covering a region of 1.7 square degrees (blue square in Fig.~\ref{fig:arc}). The observations used the VEGAS backend with the total power ON mode. The spectral resolution of OH is 358\,Hz (0.064\,km\,s$^{-1}$ at 1667\,MHz). We smoothed the data to a velocity resolution of 0.25\,km\,s$^{-1}$ to obtain a higher signal-to-noise ratio (S/N). The system temperature is $\sim$19\,K, and the beam efficiency is $\sim$71$\%$ for OH observations. The datacube was regridded to a pixel size of 3.1$'$. The total observing time is 12\,h, resulting in an rms noise level of $\sim$0.02\,K per velocity channel per pixel.

\subsection{Archive FCRAO CO (1-0) data}\label{sec:co data} The $^{12}$CO (1--0) observations were taken by the Five College Radio Astronomy Observatory (FCRAO) 13.7m telescope during the time interval from 1994 to 1997 as part of the OGS \citep[][]{Heyer1998}. The original dataset has an angular resolution of 45$''$ and a velocity resolution of 0.81\,km\,s$^{-1}$. The dataset was convolved to an angular resolution of 100.44$''$ and was corrected for the atmosphere and beam efficiency, resulting in an rms noise level of 0.16\,K per channel. The beam efficiency of FCRAO is 0.45 at 115\,GHz.

\section{Results}\label{sec:results}
\subsection{Line profiles}\label{sec:lines}

\begin{figure*}
\includegraphics[width=1.0\linewidth]{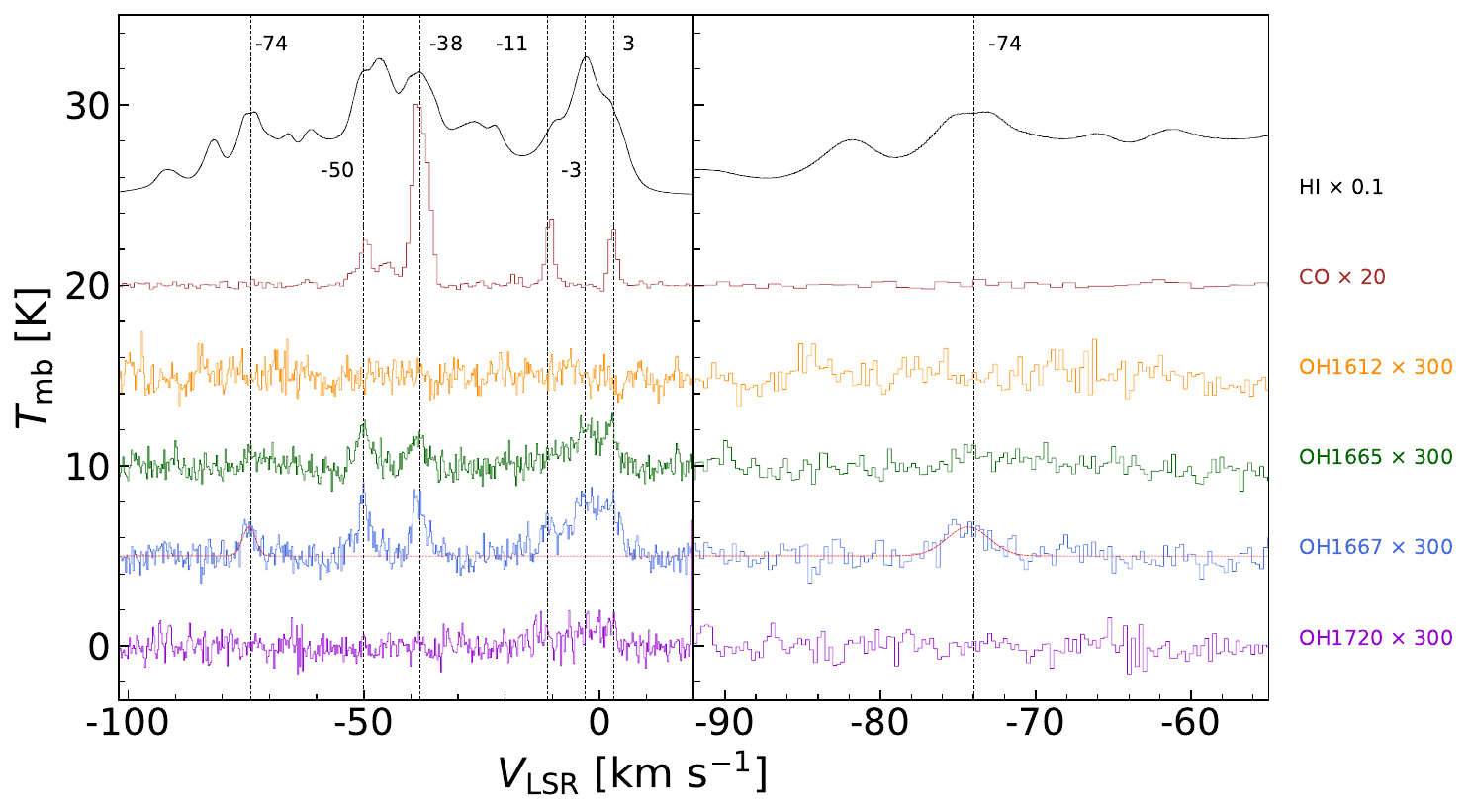}
\caption{Left: Spectra of H\,{\sc i}, CO, OH 1612, OH 1665, OH 1667, and OH 1720 MHz (from top to bottom) toward the H\,{\sc i} arc. The x-axis and y-axis denote the velocity in km\,s$^{-1}$ and the brightness temperature, respectively. Black dotted lines and velocity labels denote each of the velocity components in km\,s$^{-1}$ as seen in the OH 1667\,MHz emission. The scaling factor of each spectrum is listed in the right panel of this figure. To avoid overlapping, the spectra of H\,{\sc i}, CO, and OH have been shifted upwards by multiples of 5 K on the y-axis. Right: Same as in the left panel but showing a zoom onto the spectra at the $-$74\,km\,s$^{-1}$ component. }
\label{fig:lines} 
\end{figure*}

Previous OH observations with both the GBT and the Green Bank Observatory (GBO) 20 m telescope found that the emission of OH is weak ($T_{\rm mb} < 10$\,mK) and spatially extended toward the outer Galaxy \citep{Busch2021}. 
To obtain higher-S/N spectra, we extracted CO and OH lines from regions within a radius of 0.5$^\circ$. The resultant rms for CO, OH 1612, OH 1665, OH 1667, and OH 1720 MHz lines are 5.5, 2.4, 1.7, 1.6, and 2.0\,mK per velocity channel, respectively. Figure \ref{fig:lines} shows the spectra of H\,{\sc i}, CO, and OH. 

As can be seen, each CO velocity component has a corresponding OH 1667MHz component. However, the reverse is not always true. 
OH 1667\,MHz emission ($T_\mathrm{mb}$ = 5.4$\pm$1.4\,mK, $\Delta$\,V = 3.2$\pm$0.9\,km\,s$^{-1}$) at $-$74\,km\,s$^{-1}$ is detected without any CO emission at the 3$\sigma$ level. The particular $-$74\,km\,s$^{-1}$ velocity component is associated with the H\,{\sc i} arc structure, confirming the existence of DMG therein. 
Furthermore, a weak absorption signature at the same velocity peak as OH is identified, which is considered to be the HINSA feature originally defined in the work of \citet{Li2003}.

\subsection{Column densities}\label{sec:column density}


\subsubsection{H\,{\sc i}}\label{sec:hi}

Assuming the H\,{\sc i} 21 cm emission is optically thin,
the column density of H\,{\sc i} can be written as \citep{Draine2011}:
\begin{equation}
    N_{\rm HI} = 1.823 \times 10^{18} \int T_{\rm b}(\upsilon) d\upsilon ({\rm km\,s^{-1}}) \ [{\rm cm^{-2}}].\label{eq:hicol}
\end{equation}

The column density of H\,{\sc i} is calculated by integrating the recovered H\,{\sc i} profile ($T_\mathrm{b}(\upsilon)$) with the same velocity range as the OH emission ($-77 \sim -71$\,km\,s$^{-1}$), resulting in a value of 2.4$\times$10$^{20}$\,cm$^{-2}$. As the H\,{\sc i} profile has a much larger line width, the above value may be underestimated. Comparing with the result from Gaussian decomposition (see Appendix \ref{sec:fit
hi}), we include a 50\% uncertainty on the H\,{\sc i} column density.

\subsubsection{OH}\label{sec:oh}

The column density of OH can be written as \citep{Knapp1973, Dickey1981, Liszt1996}:
\begin{equation}
N_{\rm OH} = 2.24 \times 10^{14} \frac{T_{\rm ex}}{T_{\rm ex}-T_{\rm bg}} \int T_{1667} d\upsilon ({\rm km\,s^{-1}}) \ [{\rm cm^{-2}}],
\end{equation}
where $T_{\rm ex}$ is the excitation temperature, $T_{\rm bg}$ is the cosmic microwave background (CMB) and synchrotron emission, and $T_{1667}$ is the brightness temperature of OH 1667 MHz emission. 
The $T_{\rm ex}$ of OH was first measured from absorption against quasars, and was found to be in the range of 4-8\,K \citep{Dickey1981}.
\citet{Liszt1996} measured OH 18 cm lines through both emission and absorption profiles toward eight quasars, in which they obtained $T_\mathrm{ex}$ $\approx$ 3-5\,K. Recent work by \citet{Li2018} found that the majority ($\sim$90\%) of OH components have $T_\mathrm{ex} - T_\mathrm{bg}$ $\lesssim$ 2\,K. 
In our calculation, we adopt the value of $T_\mathrm{ex}$ - $T_\mathrm{bg}$ = 1\,K\footnote{We note that if we were to consider an extreme case that the excitation of OH is only 0.1 K above the background emission, the column density of OH would increase by a factor of 8.}. The resultant column density of OH is (2.0$\pm$0.7) $\times$ 10$^{13}$\,cm$^{-2}$. 

\subsection{HINSA}\label{sec:hinsa}

The key to obtaining the optical depth of HINSA absorption is to recover the background spectrum without the HINSA feature ($T_\mathrm{b}(\upsilon)$). The modeling of the HINSA profile follows the procedures of \citet{Krco2008}, in which we use the OH emission to set constraints on the central velocity ($\upsilon_0$) and the velocity dispersion ($\sigma$) of HINSA. 
A more detailed description of the procedures is discussed in Appendix \ref{sec:hinsa model}.

Figure \ref{fig:hinsa} shows the recovered $T_\mathrm{b}(\upsilon)$ (red dash line) and the optical depth profile of HINSA (green dotted line). The resultant central velocity is -74.1\,km\,s$^{-1}$, the full width at half maximum (FWHM) line width of HINSA is 1.53\,km\,s$^{-1}$, and the optical depth of HINSA is 0.084, corresponding to a HINSA column density of 3.8$\times$10$^{18}$ cm$^{-2}$. The uncertainty in the fitting procedures is $\sim19\%$ (see Appendix~\ref{sec:hinsa model}). 

\begin{figure}
\includegraphics[width=1.0\linewidth]{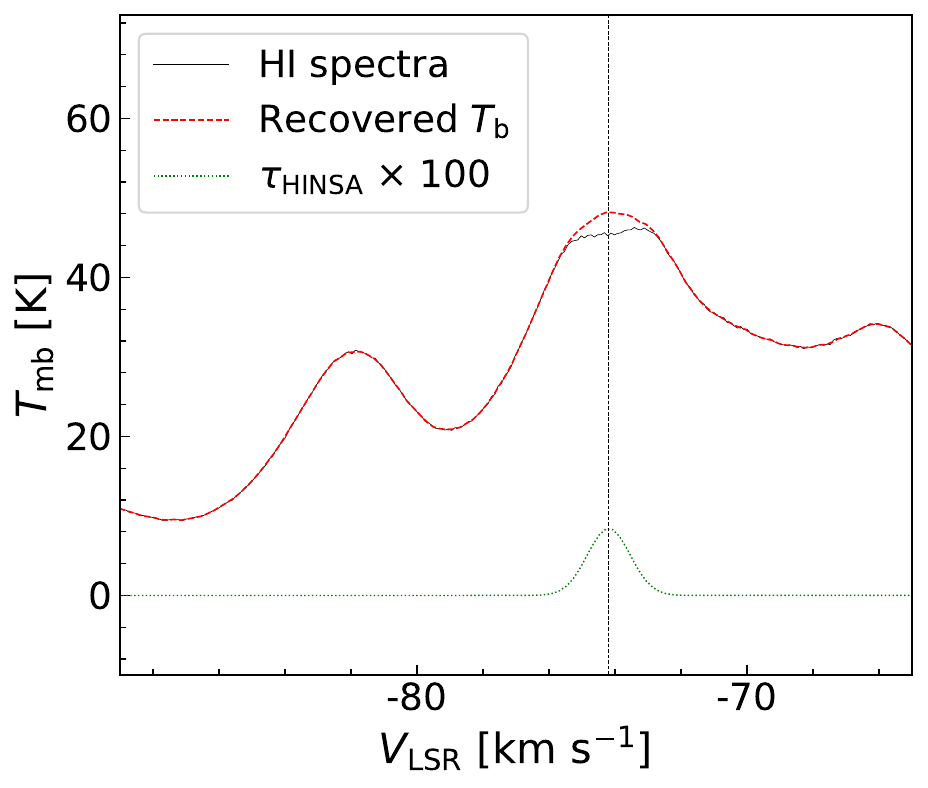}
\caption{Observed spectra of H\,{\sc i} (black solid line), recovered $T_\mathrm{b}(\upsilon)$ (red dash line), and the fitted optical depth of HINSA (green dotted line). The black vertical line represents the line velocity of HINSA.}
\label{fig:hinsa}
\end{figure}

\subsection{Molecular fraction of H\,{\sc i} arc}\label{sec:fmol}

We used the 3D $E(B-V)$ map from \citet{Green2019} to obtain the total H-nucleus column density ($N_\mathrm{H}$), in which the 3D $E(B-V)$ map is derived by combining high-quality stellar photometry from Pan-STARRS, 2MASS, and parallaxes from $Gaia$. The distance of H\,{\sc i} Arc (5.6$\pm$0.4\,kpc from the Sun) is estimated from the latest rotational curve by \citet{Reid2019}. The derived $E(B-V)$ is 0.061\,mag. Assuming $A_\mathrm{V}$/$E(B-V)=3.1$ \citep{Fitzpatrick1999}, this corresponds to $A_\mathrm{V}=0.19\,{\rm mag}$.

Observations from diffuse X-ray absorption and emission and Ly$\alpha$ absorption have shown a linear relation between the total gas column density and interstellar reddening \citep[$N_\mathrm{H}$/$E(B-V)$ = 5.8 $\times$ 10$^{21}$\,cm$^{-2}$\,mag$^{-1}$,][]{Reina1973, Bohlin1978, Predehl1995}. However, recent studies found the conversion factor to be 1.5 times higher in diffuse lines of sight (e.g., 0.015$\leq$$E(B-V)$$\leq$0.075\,mag) than the canonical value \citep{Liszt2014,Lenz2017,Nguyen2018}. As our target includes diffuse lines of sight, we adopt the value of (9.4$\pm$1.6)$\times$10$^{21}$\,cm$^{-2}$\,mag$^{-1}$ \citep{Nguyen2018}. The resultant column density of H$_2$ is $N_{\rm H_2}$ = ($N_{\rm H} - N_{\rm HI}$)/2 =(1.7$\pm$0.8)$\times$10$^{20}$\,cm$^{-2}$. On the other hand, if we assume that the abundance of OH is in the order of 10$^{-7}$ \citep{Liszt1996,Nguyen2018, Li2018}, we obtain a consistent H$_2$ column density ($N_{\rm H_2} = (2.0\pm0.7)\times10^{20}$cm$^{-2}$), as above. The inferred column density of H$_2$ is therefore over an order of magnitude higher than the previous estimate \citep[$<$10$^{19}$\,cm$^{-2}$,][]{Knee2001}.

The molecular gas fraction is given by $f_\mathrm{mol}$ =
2$N_\mathrm{H_2}$/$N_\mathrm{H}$ = 59\%$\pm$28\%, which is similar to that of nearby diffuse clouds \citep[$\sim20-44\%$,][]{Luo2020}. Our results suggest that a large fraction of molecular gas in the outer disk region of the Milky Way may be missing in the current CO surveys.

\subsection{Abundances of CO and OH}\label{sec:abundance}

The calculated abundance of OH with respect to H$_2$ is (1.2$\pm$0.7)$\times$10$^{-7}$, which is similar to the median value in nearby diffuse and translucent cloud \citep[1$\times$10$^{-7}$,][]{Liszt1996,Liszt2002, Weselak2010, Nguyen2018, Li2018, Luo2023a}. The upper limit of CO abundance is 5$\times$10$^{-7}$ (3$\sigma$, see Appendix \ref{sec:co}), which is over two orders of magnitude lower than nearby molecular clouds \citep[e.g., $\sim8-10\times10^{-5}$ in Taurus;][]{Frerking1982,Goldsmith2008}. This is consistent with high-sensitivity absorption measurements indicating that CO was only detected at regions for $A_\mathrm{V} \geq 0.32$\,mag and that the abundance of CO in diffuse clouds ranges from $2\times10^{-7}$ to $5\times10^{-6}$ \citep{Luo2020}.

\section{Discussion}\label{sec:discussion}

\subsection{The fraction of HINSA and gas volume density}\label{sec:hinsa fraction}

HINSA represents the H\,{\sc i} gas that resides in the interior of the molecular cloud, which is produced by the cosmic ray(CR)-induced dissociation of H$_2$. Thus, in chemical equilibrium, the steady-state HINSA density mainly depends on the CR ionization rate ($\zeta_2$), which is given by \citep{Goldsmith2005}:

\begin{equation}
    n_{\rm HI} = \frac{\zeta_2}{2k'},
\end{equation}
where k$'$ = $4\times10^{-17}$\,cm$^3$\,s$^{-1}$ is the H$_2$ formation rate coefficient \citep{Gry2002}. Assuming that $\zeta_2$ decreases exponentially with the increasing $R_{\rm GC}$ \citep{Wolfire2003}, we adopt a value of $1.6 \times 10^{-16}$\,s$^{-1}$, which is five times smaller than in nearby diffuse clouds of similar extinction \citep{Luo2023b}. Thus, the steady-state density of H\,{\sc i} is $n_\mathrm{HI}$ = 2.0\,cm$^{-3}$.

The fractional abundance of HINSA ($N_{\rm HINSA}$/$N_{\rm H_2}$) in DMG is 0.022$\pm$0.011, which is over an order of magnitude higher than that in nearby clouds \citep[10$^{-3}$,][]{Li2003,Krco2010,Tang2021a}, but is similar to the value in the early evolutionary stage of dark molecular clouds \citep[0.02,][]{Zuo2018}. The inferred molecular gas density at steady state is $n_{\rm H_2} = 91\pm46$\,cm$^{-3}$, which is similar to the measured density in nearby diffuse clouds \citep{Luo2020,Luo2023b}.

\subsection{The formation mechanism of the DMG in the H\,{\sc i} arc}\label{sec:formation}

The shell-like structure is a common feature in both Milky Way and external galaxies \citep{Heiles1979,Tenorio-Tagle1988,Hatzidimitriou2005,Ehlerova2005,Dawson2013,Ehlerova2016}. The enormous H\,{\sc i} supershells can be driven by the powerful stellar wind and radiation from massive stars and star clusters, and supernova explosions \citep{McCray1987,Elmegreen1998,Seifried2020}. In the solar neighborhood, giant molecular clouds are found to be at the surface of the Local Bubble, which is thought to have been induced by a supernova explosion $\sim$14 Myr ago \citep{Zucker2022}. 

The existence of DMG in the cold H\,{\sc i} arc structure and the corresponding large H\,{\sc i} supershell GSH139-03-69 suggests a plausible scenario in which the formation of molecular gas is triggered by the expansion of the supershell. The expansion of the supershell generates powerful shockwaves that sweep up and compress the surrounding low-density ISM, leading to an enhancement of gas density as well as gravitational instability \citep{Woodward1976,Bisnovatyi-Kogan1995,Elmegreen1998,Wunsch2010}. 

The crossing timescale over which the shock waves pass through the H\,{\sc i} arc is on the order of 10$^7$ yr \citep[assuming a typical shock velocity of 10\,km\,s$^{-1}$,][]{McCray1987,Inutsuka2015}. This timescale is much longer than the lifetime of giant molecular clouds (a few million years) and is also longer than the timescale required for H\,{\sc i} to be converted into H$_2$ ($t_{\rm HI-to-H_2} = 1/(2k'n_{\rm HI}+\zeta_2) \approx 10^6$ yr), enabling the formation of molecular gas during the expansion of the H\,{\sc i} supershell. However, the low gas density and extinction may still be insufficient to build up the abundance of CO \citep{Inutsuka2015,Seifried2020}, which causes the H\,{\sc i} arc to be dominated by DMG. Also, because the timescale on which the shell collapses to trigger star formation is inversely proportional to the midplane density \citep{Elmegreen2002}, supershell-driven cloud and star formation of this kind could be more frequent in the outer regions than in the nuclear regions of the galaxy. Our photodissociation regions (PDR) simulations suggest a cloud density of between 10$^{2}$ and 10$^{2.5}$\,cm$^{-3}$ and an evolution timescale of between 10$^5$ and 10$^{6}$\,yr, which is consistent with the above analysis (see Appendix \ref{sec:pdr}). 

For a rough estimate, assuming the cloud has a typical scale of a few to tens of parsecs, the DMG mass is on the order of $10^4 \sim 10^5$ M$_\odot$. Given the large number of supershells identified in the anticenter direction \citep[a total number of 566;][]{Suad2014}, the DMG mass that is associated with the supershells could be on the order of $10^7 \sim 10^8$ M$_\odot$. This is of the same order of magnitude as the estimation of CO-dark diffuse molecular mass in the outer Galaxy \citep{Busch2021} and the total mass traced by CO in the outer Galaxy \cite[$\sim10^8$ M$_\odot$,][]{Heyer2015}, suggesting that half of the molecular mass could be missed in CO surveys.

\section{Conclusion}\label{sec:conclusion}
We confirm the existence of DMG toward a cold H\,{\sc i} arc. This structure is not supposed to exist according to the canonical phased ISM models. The analysis from H\,{\sc i} and OH spectra suggests an early evolution stage ($t \lesssim 10^6$ yr) of low-density DMG. Our main conclusions are as follows:

\begin{enumerate}
    \item  We detect OH emission and the signature of HINSA toward the cold H\,{\sc i} arc, where no CO emission is found. The existence of DMG associated with the H\,{\sc i} arc supports the scenario that the expansion of the H\,{\sc i} supershell triggers the formation of the molecular cloud. 
    
    \item The inferred H$_2$ column density (1.7$\pm$0.8$\times$10$^{20}$\,cm$^{-2}$) is over an order of magnitude higher than the previous estimate. The resultant fractional abundance of HINSA is 0.022$\pm$0.011, which is an order of magnitude higher than that of nearby evolved clouds, but is similar to the values in the early evolutionary stage of dark clouds.
    
    \item The derived column density of OH is (2.0$\pm$0.7)$\times$10$^{13}$\,cm$^{-2}$, and the resultant abundance is 1.2$\pm$0.7$\times$10$^{-7}$. The upper limit of CO abundance is 5$\times$10$^{-7}$, which is more than an order of magnitude lower than that in nearby clouds.

    \item The visual extinction ($A_{\rm V}$ = 0.19$\pm$0.03\,mag), the inferred molecular fraction (59\%$\pm$28\%), and the gas density ($n_{\rm H_2}$ = $91\pm46$\,cm$^{-3}$) of the DMG are similar to those of nearby diffuse clouds, suggesting similar properties of the CO-dark gas between the local and outer disk.

\end{enumerate}

In addition, our findings show that the distribution of molecular gas in the outer disk regions could be much more extended than the gas traced by CO. This is consistent with previous high-sensitivity OH observations, where a diffuse thick DMG disk is traced by weak extended OH emission in the outer Galaxy \citep{Busch2021}. Future large OH surveys and an array covering the OH lines on FAST would improve our understanding of DMG in the outer Galaxy.

\begin{acknowledgements}
We are grateful to the referee for the insightful review of our manuscript, which improved the quality and clarity of our work.
We thank the scientific staff and telescope operators at the Green Bank Observatory (GBO) for their advice and assistance with the remote observations, in particular Tapasi Ghosh, and for the support of the data reduction, especially Pedro Salas.

This work has been supported by the National Natural Science Foundation of China (grant No. 12041305,12173016), Cultivation Project for FAST scientific Payoff and Research achievement of CAMS-CAS, China Postdoctoral Science Foundation (grant No. 2021M691533), the Program for Innovative Talents, Entrepreneur in Jiangsu, the science research grants from the China Manned Space Project with NO.CMS-CSST-2021-A08. N.Y.T. acknowledges support from the University Annual Scientific Research Plan of Anhui Province (No. 2023AH030052, No. 2022AH010013), Zhejiang Lab Open Research Project (NO. K2022PE0AB01), Cultivation Project for FAST Scientific Payoff and Research Achievement of CAMS-CAS.

The Green Bank Observatory is a facility of the National Science Foundation operated under cooperative agreement by Associated Universities, Inc. The Canadian Galactic Plane Survey (CGPS) is a Canadian project with international partners. The Dominion Radio Astrophysical Observatory is operated as a national facility by the National Research Council of Canada. The Five College Radio Astronomy Observatory CO Survey of the Outer Galaxy was supported by NSF grant AST 94-20159. The CGPS is supported by a grant from the Natural Sciences and Engineering Research Council of Canada.
\end{acknowledgements}


%
%
\bibliographystyle{aa}
\bibliography{reference} 

\begin{thebibliography}{77}
\expandafter\ifx\csname natexlab\endcsname\relax\def\natexlab#1{#1}\fi

\bibitem[{{Bisbas} {et~al.}(2012){Bisbas}, {Bell}, {Viti}, {Yates}, \&
  {Barlow}}]{Bisbas2012}
{Bisbas}, T.~G., {Bell}, T.~A., {Viti}, S., {Yates}, J., \& {Barlow}, M.~J.
  2012, \mnras, 427, 2100

\bibitem[{{Bisbas} {et~al.}(2019){Bisbas}, {Schruba}, \& {van
  Dishoeck}}]{Bisbas2019}
{Bisbas}, T.~G., {Schruba}, A., \& {van Dishoeck}, E.~F. 2019, \mnras, 485,
  3097

\bibitem[{{Bisnovatyi-Kogan} \& {Silich}(1995)}]{Bisnovatyi-Kogan1995}
{Bisnovatyi-Kogan}, G.~S. \& {Silich}, S.~A. 1995, Reviews of Modern Physics,
  67, 661

\bibitem[{{Bohlin} {et~al.}(1978){Bohlin}, {Savage}, \& {Drake}}]{Bohlin1978}
{Bohlin}, R.~C., {Savage}, B.~D., \& {Drake}, J.~F. 1978, \apj, 224, 132

\bibitem[{{Bolatto} {et~al.}(2013){Bolatto}, {Wolfire}, \&
  {Leroy}}]{Bolatto2013}
{Bolatto}, A.~D., {Wolfire}, M., \& {Leroy}, A.~K. 2013, \araa, 51, 207

\bibitem[{{Busch} {et~al.}(2021){Busch}, {Engelke}, {Allen}, \&
  {Hogg}}]{Busch2021}
{Busch}, M.~P., {Engelke}, P.~D., {Allen}, R.~J., \& {Hogg}, D.~E. 2021, \apj,
  914, 72

\bibitem[{{Dame} {et~al.}(2001){Dame}, {Hartmann}, \& {Thaddeus}}]{Dame2001}
{Dame}, T.~M., {Hartmann}, D., \& {Thaddeus}, P. 2001, \apj, 547, 792

\bibitem[{{Dawson} {et~al.}(2013){Dawson}, {McClure-Griffiths}, {Wong},
  {Dickey}, {Hughes}, {Fukui}, \& {Kawamura}}]{Dawson2013}
{Dawson}, J.~R., {McClure-Griffiths}, N.~M., {Wong}, T., {et~al.} 2013, \apj,
  763, 56

\bibitem[{{Dickey} {et~al.}(1981){Dickey}, {Crovisier}, \&
  {Kazes}}]{Dickey1981}
{Dickey}, J.~M., {Crovisier}, J., \& {Kazes}, I. 1981, \aap, 98, 271

\bibitem[{{Dickey} \& {Lockman}(1990)}]{Dickey1990}
{Dickey}, J.~M. \& {Lockman}, F.~J. 1990, \araa, 28, 215

\bibitem[{{Draine}(1978)}]{Draine1978}
{Draine}, B.~T. 1978, \apjs, 36, 595

\bibitem[{{Draine}(2011)}]{Draine2011}
{Draine}, B.~T. 2011, {Physics of the Interstellar and Intergalactic Medium}

\bibitem[{{Ehlerov{\'a}} \& {Palou{\v{s}}}(2005)}]{Ehlerova2005}
{Ehlerov{\'a}}, S. \& {Palou{\v{s}}}, J. 2005, \aap, 437, 101

\bibitem[{{Ehlerov{\'a}} \& {Palou{\v{s}}}(2016)}]{Ehlerova2016}
{Ehlerov{\'a}}, S. \& {Palou{\v{s}}}, J. 2016, \aap, 587, A5

\bibitem[{{Elmegreen}(1998)}]{Elmegreen1998}
{Elmegreen}, B.~G. 1998, in Astronomical Society of the Pacific Conference
  Series, Vol. 148, Origins, ed. C.~E. {Woodward}, J.~M. {Shull}, \&
  J.~{Thronson}, Harley~A., 150

\bibitem[{{Elmegreen} {et~al.}(2002){Elmegreen}, {Palou{\v{s}}}, \&
  {Ehlerov{\'a}}}]{Elmegreen2002}
{Elmegreen}, B.~G., {Palou{\v{s}}}, J., \& {Ehlerov{\'a}}, S. 2002, \mnras,
  334, 693

\bibitem[{{Federman} {et~al.}(1996){Federman}, {Weber}, \&
  {Lambert}}]{Federman1996}
{Federman}, S.~R., {Weber}, J., \& {Lambert}, D.~L. 1996, \apj, 463, 181

\bibitem[{{Field} {et~al.}(1969){Field}, {Goldsmith}, \& {Habing}}]{Field1969}
{Field}, G.~B., {Goldsmith}, D.~W., \& {Habing}, H.~J. 1969, \apjl, 155, L149

\bibitem[{{Fitzpatrick}(1999)}]{Fitzpatrick1999}
{Fitzpatrick}, E.~L. 1999, \pasp, 111, 63

\bibitem[{{Frerking} {et~al.}(1982){Frerking}, {Langer}, \&
  {Wilson}}]{Frerking1982}
{Frerking}, M.~A., {Langer}, W.~D., \& {Wilson}, R.~W. 1982, \apj, 262, 590

\bibitem[{{Goldsmith}(2013)}]{Goldsmith2013}
{Goldsmith}, P.~F. 2013, \apj, 774, 134

\bibitem[{{Goldsmith} {et~al.}(2008){Goldsmith}, {Heyer}, {Narayanan}, {Snell},
  {Li}, \& {Brunt}}]{Goldsmith2008}
{Goldsmith}, P.~F., {Heyer}, M., {Narayanan}, G., {et~al.} 2008, \apj, 680, 428

\bibitem[{{Goldsmith} \& {Li}(2005)}]{Goldsmith2005}
{Goldsmith}, P.~F. \& {Li}, D. 2005, \apj, 622, 938

\bibitem[{{Green} {et~al.}(2019){Green}, {Schlafly}, {Zucker}, {Speagle}, \&
  {Finkbeiner}}]{Green2019}
{Green}, G.~M., {Schlafly}, E., {Zucker}, C., {Speagle}, J.~S., \&
  {Finkbeiner}, D. 2019, \apj, 887, 93

\bibitem[{{Grenier} {et~al.}(2005){Grenier}, {Casandjian}, \&
  {Terrier}}]{Grenier2005}
{Grenier}, I.~A., {Casandjian}, J.-M., \& {Terrier}, R. 2005, Science, 307,
  1292

\bibitem[{{Gry} {et~al.}(2002){Gry}, {Boulanger}, {Nehm{\'e}}, {Pineau des
  For{\^e}ts}, {Habart}, \& {Falgarone}}]{Gry2002}
{Gry}, C., {Boulanger}, F., {Nehm{\'e}}, C., {et~al.} 2002, \aap, 391, 675

\bibitem[{{Haslam} {et~al.}(1982){Haslam}, {Salter}, {Stoffel}, \&
  {Wilson}}]{Haslam1982}
{Haslam}, C.~G.~T., {Salter}, C.~J., {Stoffel}, H., \& {Wilson}, W.~E. 1982,
  \aaps, 47, 1

\bibitem[{{Hatzidimitriou} {et~al.}(2005){Hatzidimitriou}, {Stanimirovic},
  {Maragoudaki}, {Staveley-Smith}, {Dapergolas}, \&
  {Bratsolis}}]{Hatzidimitriou2005}
{Hatzidimitriou}, D., {Stanimirovic}, S., {Maragoudaki}, F., {et~al.} 2005,
  \mnras, 360, 1171

\bibitem[{{Heiles}(1979)}]{Heiles1979}
{Heiles}, C. 1979, \apj, 229, 533

\bibitem[{{Heiles} \& {Troland}(2003)}]{Heiles2003}
{Heiles}, C. \& {Troland}, T.~H. 2003, \apj, 586, 1067

\bibitem[{{Heyer} \& {Dame}(2015)}]{Heyer2015}
{Heyer}, M. \& {Dame}, T.~M. 2015, \araa, 53, 583

\bibitem[{{Heyer} {et~al.}(1998){Heyer}, {Brunt}, {Snell}, {Howe}, {Schloerb},
  \& {Carpenter}}]{Heyer1998}
{Heyer}, M.~H., {Brunt}, C., {Snell}, R.~L., {et~al.} 1998, \apjs, 115, 241

\bibitem[{{Hu} {et~al.}(2021){Hu}, {Sternberg}, \& {van Dishoeck}}]{Hu2021}
{Hu}, C.-Y., {Sternberg}, A., \& {van Dishoeck}, E.~F. 2021, \apj, 920, 44

\bibitem[{{Inutsuka} {et~al.}(2015){Inutsuka}, {Inoue}, {Iwasaki}, \&
  {Hosokawa}}]{Inutsuka2015}
{Inutsuka}, S.-i., {Inoue}, T., {Iwasaki}, K., \& {Hosokawa}, T. 2015, \aap,
  580, A49

\bibitem[{{Kalberla} \& {Kerp}(2009)}]{Kalberla2009}
{Kalberla}, P. M.~W. \& {Kerp}, J. 2009, \araa, 47, 27

\bibitem[{{Knapp} \& {Kerr}(1973)}]{Knapp1973}
{Knapp}, G.~R. \& {Kerr}, F.~J. 1973, \aj, 78, 453

\bibitem[{{Knee} \& {Brunt}(2001)}]{Knee2001}
{Knee}, L. B.~G. \& {Brunt}, C.~M. 2001, \nat, 412, 308

\bibitem[{{Kr{\v{c}}o} \& {Goldsmith}(2010)}]{Krco2010}
{Kr{\v{c}}o}, M. \& {Goldsmith}, P.~F. 2010, \apj, 724, 1402

\bibitem[{{Kr{\v{c}}o} {et~al.}(2008){Kr{\v{c}}o}, {Goldsmith}, {Brown}, \&
  {Li}}]{Krco2008}
{Kr{\v{c}}o}, M., {Goldsmith}, P.~F., {Brown}, R.~L., \& {Li}, D. 2008, \apj,
  689, 276

\bibitem[{{Lee} {et~al.}(2015){Lee}, {Stanimirovi{\'c}}, {Murray}, {Heiles}, \&
  {Miller}}]{Lee2015}
{Lee}, M.-Y., {Stanimirovi{\'c}}, S., {Murray}, C.~E., {Heiles}, C., \&
  {Miller}, J. 2015, \apj, 809, 56

\bibitem[{{Lenz} {et~al.}(2017){Lenz}, {Hensley}, \& {Dor{\'e}}}]{Lenz2017}
{Lenz}, D., {Hensley}, B.~S., \& {Dor{\'e}}, O. 2017, \apj, 846, 38

\bibitem[{{Li} \& {Goldsmith}(2003)}]{Li2003}
{Li}, D. \& {Goldsmith}, P.~F. 2003, \apj, 585, 823

\bibitem[{{Li} {et~al.}(2018){Li}, {Tang}, {Nguyen}, {Dawson}, {Heiles}, {Xu},
  {Pan}, {Goldsmith}, {Gibson}, {Murray}, {Robishaw}, {McClure-Griffiths},
  {Dickey}, {Pineda}, {Stanimirovi{\'c}}, {Bronfman}, {Troland}, \& {PRIMO
  Collaboration}}]{Li2018}
{Li}, D., {Tang}, N., {Nguyen}, H., {et~al.} 2018, \apjs, 235, 1

\bibitem[{{Liszt}(2014)}]{Liszt2014}
{Liszt}, H. 2014, \apj, 780, 10

\bibitem[{{Liszt} \& {Lucas}(1996)}]{Liszt1996}
{Liszt}, H. \& {Lucas}, R. 1996, \aap, 314, 917

\bibitem[{{Liszt} \& {Lucas}(2002)}]{Liszt2002}
{Liszt}, H. \& {Lucas}, R. 2002, \aap, 391, 693

\bibitem[{{Luo} {et~al.}(2020){Luo}, {Li}, {Tang}, {Dawson}, {Dickey},
  {Bronfman}, {Qin}, {Gibson}, {Plambeck}, {Finger}, {Green}, {Mardones},
  {Koo}, \& {Lo}}]{Luo2020}
{Luo}, G., {Li}, D., {Tang}, N., {et~al.} 2020, \apjl, 889, L4

\bibitem[{{Luo} {et~al.}(2023{\natexlab{a}}){Luo}, {Zhang}, {Bisbas}, {Li},
  {Tang}, {Wang}, {Zhou}, {Zuo}, {Yue}, {Zhou}, \& {Lin}}]{Luo2023a}
{Luo}, G., {Zhang}, Z.-Y., {Bisbas}, T.~G., {et~al.} 2023{\natexlab{a}}, \apj,
  942, 101

\bibitem[{{Luo} {et~al.}(2023{\natexlab{b}}){Luo}, {Zhang}, {Bisbas}, {Li},
  {Zhou}, {Tang}, {Wang}, {Zuo}, \& {Yue}}]{Luo2023b}
{Luo}, G., {Zhang}, Z.-Y., {Bisbas}, T.~G., {et~al.} 2023{\natexlab{b}}, \apj,
  946, 91

\bibitem[{{Mangum} \& {Shirley}(2015)}]{Mangum2015}
{Mangum}, J.~G. \& {Shirley}, Y.~L. 2015, \pasp, 127, 266

\bibitem[{{McCray} \& {Kafatos}(1987)}]{McCray1987}
{McCray}, R. \& {Kafatos}, M. 1987, \apj, 317, 190

\bibitem[{{McElroy} {et~al.}(2013){McElroy}, {Walsh}, {Markwick}, {Cordiner},
  {Smith}, \& {Millar}}]{McElroy2013}
{McElroy}, D., {Walsh}, C., {Markwick}, A.~J., {et~al.} 2013, \aap, 550, A36

\bibitem[{{McKee} \& {Ostriker}(1977)}]{McKee1977}
{McKee}, C.~F. \& {Ostriker}, J.~P. 1977, \apj, 218, 148

\bibitem[{{M{\'e}ndez-Delgado} {et~al.}(2022){M{\'e}ndez-Delgado}, {Amayo},
  {Arellano-C{\'o}rdova}, {Esteban}, {Garc{\'\i}a-Rojas}, {Carigi}, \&
  {Delgado-Inglada}}]{Mendez-Delgado2022}
{M{\'e}ndez-Delgado}, J.~E., {Amayo}, A., {Arellano-C{\'o}rdova}, K.~Z.,
  {et~al.} 2022, \mnras, 510, 4436

\bibitem[{{Nguyen} {et~al.}(2018){Nguyen}, {Dawson}, {Miville-Desch{\^e}nes},
  {Tang}, {Li}, {Heiles}, {Murray}, {Stanimirovi{\'c}}, {Gibson},
  {McClure-Griffiths}, {Troland}, {Bronfman}, \& {Finger}}]{Nguyen2018}
{Nguyen}, H., {Dawson}, J.~R., {Miville-Desch{\^e}nes}, M.~A., {et~al.} 2018,
  \apj, 862, 49

\bibitem[{{Pineda} {et~al.}(2013){Pineda}, {Langer}, {Velusamy}, \&
  {Goldsmith}}]{Pineda2013}
{Pineda}, J.~L., {Langer}, W.~D., {Velusamy}, T., \& {Goldsmith}, P.~F. 2013,
  \aap, 554, A103

\bibitem[{{Planck Collaboration} {et~al.}(2011){Planck Collaboration}, {Ade},
  {Aghanim}, {Arnaud}, {Ashdown}, {Aumont}, {Baccigalupi}, {Balbi}, {Banday},
  {Barreiro}, \& et~al.}]{Planck2011}
{Planck Collaboration}, {Ade}, P.~A.~R., {Aghanim}, N., {et~al.} 2011, \aap,
  536, A19

\bibitem[{{Predehl} \& {Schmitt}(1995)}]{Predehl1995}
{Predehl}, P. \& {Schmitt}, J.~H.~M.~M. 1995, \aap, 500, 459

\bibitem[{{Reid} {et~al.}(2019){Reid}, {Menten}, {Brunthaler}, {Zheng}, {Dame},
  {Xu}, {Li}, {Sakai}, {Wu}, {Immer}, {Zhang}, {Sanna}, {Moscadelli}, {Rygl},
  {Bartkiewicz}, {Hu}, {Quiroga-Nu{\~n}ez}, \& {van Langevelde}}]{Reid2019}
{Reid}, M.~J., {Menten}, K.~M., {Brunthaler}, A., {et~al.} 2019, \apj, 885, 131

\bibitem[{{Reina} \& {Tarenghi}(1973)}]{Reina1973}
{Reina}, C. \& {Tarenghi}, M. 1973, \aap, 26, 257

\bibitem[{{Remy} {et~al.}(2017){Remy}, {Grenier}, {Marshall}, \&
  {Casandjian}}]{Remy2017}
{Remy}, Q., {Grenier}, I.~A., {Marshall}, D.~J., \& {Casandjian}, J.~M. 2017,
  \aap, 601, A78

\bibitem[{{Rybarczyk} {et~al.}(2022){Rybarczyk}, {Stanimirovi{\'c}}, {Gong},
  {Babler}, {Murray}, {Gerin}, {Winters}, {Luo}, {Dame}, \&
  {Steffes}}]{Rybarczyk2022}
{Rybarczyk}, D.~R., {Stanimirovi{\'c}}, S., {Gong}, M., {et~al.} 2022, \apj,
  928, 79

\bibitem[{{Sahai} \& {Nyman}(1997)}]{Sahai1997}
{Sahai}, R. \& {Nyman}, L.-{\r{A}}. 1997, \apjl, 487, L155

\bibitem[{{Sahai} {et~al.}(2013){Sahai}, {Vlemmings}, {Huggins}, {Nyman}, \&
  {Gonidakis}}]{Sahai2013}
{Sahai}, R., {Vlemmings}, W.~H.~T., {Huggins}, P.~J., {Nyman}, L.~{\r{A}}., \&
  {Gonidakis}, I. 2013, \apj, 777, 92

\bibitem[{{Seifried} {et~al.}(2020){Seifried}, {Haid}, {Walch}, {Borchert}, \&
  {Bisbas}}]{Seifried2020}
{Seifried}, D., {Haid}, S., {Walch}, S., {Borchert}, E.~M.~A., \& {Bisbas},
  T.~G. 2020, \mnras, 492, 1465

\bibitem[{{Suad} {et~al.}(2014){Suad}, {Caiafa}, {Arnal}, \&
  {Cichowolski}}]{Suad2014}
{Suad}, L.~A., {Caiafa}, C.~F., {Arnal}, E.~M., \& {Cichowolski}, S. 2014,
  \aap, 564, A116

\bibitem[{{Tang} {et~al.}(2021){Tang}, {Li}, {Yue}, {Zuo}, {Liu}, {Luo},
  {Chen}, {Qin}, {Wu}, \& {Heiles}}]{Tang2021a}
{Tang}, N., {Li}, D., {Yue}, N., {et~al.} 2021, \apjs, 252, 1

\bibitem[{{Tang} {et~al.}(2020){Tang}, {Zuo}, {Li}, {Qian}, {Liu}, {Wu},
  {Kr{\v{c}}o}, {Liu}, {Yue}, {Zhu}, {Liu}, {Yu}, {Sun}, {Jiang}, {Pan}, {Li},
  {Gan}, {Yao}, {Liu}, \& {FAST Collaboration}}]{2020RAA....20...77T}
{Tang}, N.-Y., {Zuo}, P., {Li}, D., {et~al.} 2020, Research in Astronomy and
  Astrophysics, 20, 077

\bibitem[{{Taylor} {et~al.}(2003){Taylor}, {Gibson}, {Peracaula}, {Martin},
  {Landecker}, {Brunt}, {Dewdney}, {Dougherty}, {Gray}, {Higgs}, {Kerton},
  {Knee}, {Kothes}, {Purton}, {Uyaniker}, {Wallace}, {Willis}, \&
  {Durand}}]{Taylor2003}
{Taylor}, A.~R., {Gibson}, S.~J., {Peracaula}, M., {et~al.} 2003, \aj, 125,
  3145

\bibitem[{{Tenorio-Tagle} \& {Bodenheimer}(1988)}]{Tenorio-Tagle1988}
{Tenorio-Tagle}, G. \& {Bodenheimer}, P. 1988, \araa, 26, 145

\bibitem[{{Weselak} {et~al.}(2010){Weselak}, {Galazutdinov}, {Beletsky}, \&
  {Kre{\l}owski}}]{Weselak2010}
{Weselak}, T., {Galazutdinov}, G.~A., {Beletsky}, Y., \& {Kre{\l}owski}, J.
  2010, \mnras, 402, 1991

\bibitem[{{Wilson} {et~al.}(1970){Wilson}, {Jefferts}, \&
  {Penzias}}]{Wilson1970}
{Wilson}, R.~W., {Jefferts}, K.~B., \& {Penzias}, A.~A. 1970, \apjl, 161, L43

\bibitem[{{Wolfire} {et~al.}(2003){Wolfire}, {McKee}, {Hollenbach}, \&
  {Tielens}}]{Wolfire2003}
{Wolfire}, M.~G., {McKee}, C.~F., {Hollenbach}, D., \& {Tielens}, A.~G.~G.~M.
  2003, \apj, 587, 278

\bibitem[{{Woodward}(1976)}]{Woodward1976}
{Woodward}, P.~R. 1976, \apj, 207, 484

\bibitem[{{W{\"u}nsch} {et~al.}(2010){W{\"u}nsch}, {Dale}, {Palou{\v{s}}}, \&
  {Whitworth}}]{Wunsch2010}
{W{\"u}nsch}, R., {Dale}, J.~E., {Palou{\v{s}}}, J., \& {Whitworth}, A.~P.
  2010, \mnras, 407, 1963

\bibitem[{{Zucker} {et~al.}(2022){Zucker}, {Goodman}, {Alves}, {Bialy},
  {Foley}, {Speagle}, {Gro{\^I}{\texttwosuperior}schedl}, {Finkbeiner},
  {Burkert}, {Khimey}, \& {Swiggum}}]{Zucker2022}
{Zucker}, C., {Goodman}, A.~A., {Alves}, J., {et~al.} 2022, \nat, 601, 334

\bibitem[{{Zuo} {et~al.}(2018){Zuo}, {Li}, {Peek}, {Chang}, {Zhang}, {Chapman},
  {Goldsmith}, \& {Zhang}}]{Zuo2018}
{Zuo}, P., {Li}, D., {Peek}, J.~E.~G., {et~al.} 2018, \apj, 867, 13

\end{thebibliography}


\begin{appendix} 

\section{Gaussian decomposition of H\,{\sc i}}\label{sec:fit hi}

\begin{figure}
\includegraphics[width=1.0\linewidth]{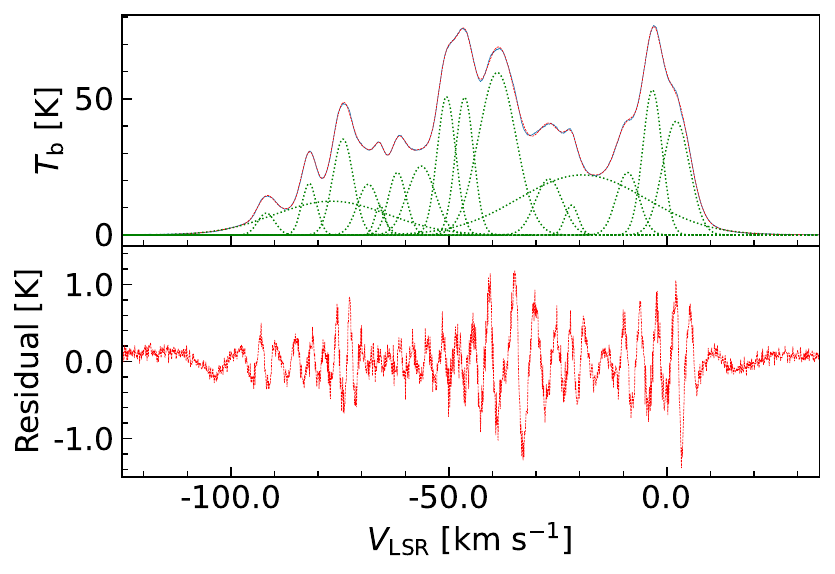}
\caption{Gaussian decomposition of the recovered $T_{\rm b}(\upsilon)$ (upper panel) and the residual after the fitting (lower panel).}
\label{fig:higs}
\end{figure}

The H\,{\sc i} profile usually exhibits a larger line width than the molecular line counterpart. To evaluate the uncertainties in deriving the H\,{\sc i} column density in Section \ref{sec:hi}, we performed Gaussian decomposition of the H\,{\sc i} profile. Figure \ref{fig:higs} shows the Gaussian decomposition of the recovered $T_{\rm b}(\upsilon)$ profiles and the residual after the fitting. The Gaussian fitting result at $-$74.1\,km\,s$^{-1}$ has a brightness temperature of 35$\pm$1 K and a line width of 5.5$\pm$0.1 km\,s$^{-1}$. The H\,{\sc i} column density calculated from equation \ref{eq:hicol} is (3.7$\pm$0.1)$\times$10$^{20}$\,cm$^{-2}$, which is 50\% higher than the value in Section \ref{sec:hi}. Thus, we consider that there is an overall 50\% uncertainty on the H\,{\sc i} column density.

\section{modeling of the HINSA profile}\label{sec:hinsa model}

Considering the three-component H\,{\sc i} model, the observed H\,{\sc i} profile ($T_\mathrm{R}$) is a combination of background emission, cold H\,{\sc i} absorption against the background, and foreground emission between the cold H\,{\sc i} and the observer. Assuming that the atomic cloud has a uniform temperature and small optical depth, following equation (8) in \citet{Li2003}, the radiative transfer equation of the modeled background spectrum can be written as:
\begin{equation}
    T_{\rm{b}}(\upsilon) = \frac{T_{\rm{R}} + (T_{\rm{c}} - T_{\rm{k}})(1 - \tau_{\rm{f}})(1 - e^{-\tau(\upsilon)})}{1 - p(1-e^{-\tau(\upsilon)})},
\end{equation}
where $T_\mathrm{R}$ is the observed spectrum; $T_\mathrm{c}$ is the background continuum, which includes the cosmic microwave background (CMB, 2.73\,K) and the Galactic synchrotron background; $T_\mathrm{k}$ is the excitation temperature of the cold H\,{\sc i}, which is equal to the kinetic temperature of molecular gas (here we take a typical kinetic temperature of 15\,K); $\tau_\mathrm{f}$ and $\tau(\upsilon)$ are the optical depth of the foreground H\,{\sc i} gas and HINSA, respectively; $p$ is the fraction of the H\,{\sc i} in the background, which is defined as $p = \tau_\mathrm{b}/\tau_\mathrm{tot}$;
$\tau_\mathrm{f}$ is assumed as 0.1 in our calculation, which would only result in $<$3\% uncertainty on the optical depth of HINSA as long as the optically thin assumption is valid; and
$T_\mathrm{c}$ is estimated from the 408\,MHz all-sky continuum survey through the equation: $T_\mathrm{c} = T_\mathrm{CMB} + T_{408}(\nu/408)^{-2.8}$ \citep{Haslam1982}. The continuum emission in our target region is 60\,K at 408\,MHz, resulting in a $T_\mathrm{c}$ of 4.6\,K.

Following equation (9) in \citet{Li2003}, $p$ can be estimated from the geometry of the galactic disk:
\begin{equation}
    p = {\rm{erfc}}[\frac{\sqrt{4 ln2} D sin(b)}{z}],
\end{equation}
where D is the distance of the cloud, b is the Galactic latitude, and z is the scale height. With the full consideration of H\,{\sc i} density distribution, the kinematic distance, and the scale height at the cloud position. In our calculations, we take a $p$ value of $\sim$0.9. We stress here that the $p$ value should not be much less than 1; otherwise, the HINSA signature would be obscured by the foreground emission \citep{Krco2008}. 

To recover the background spectrum without HINSA, we follow the procedures in \citet{Krco2008}. The optical depth of HINSA ($\tau(\upsilon)$) can be described as a Gaussian function:
\begin{equation}
    \tau(\upsilon) = \tau_0 e^{-\frac{(\upsilon-\upsilon_0)^2}{2\sigma^2}},
\end{equation}
where $\tau_0$ is the peak of optical depth, $\upsilon_0$ is the central velocity, and $\sigma$ is the velocity dispersion of HINSA. We used the OH emission to set constraints on the central velocity ($\upsilon_0$) and velocity dispersion ($\sigma$) of HINSA.

As the second derivative of a Gaussian function is sensitive to the velocity dispersion, and the velocity dispersion of HINSA ($\sim$1\,km\,s$^{-1}$) is much narrower than that of background H\,{\sc i} ($>5$\,km\,s$^{-1}$), the second derivative of H\,{\sc i} profile usually have large jumps at the velocity range of HINSA, even when the HINSA is relatively weak \citep{Krco2008}. 
Thus, minimizing the integration of the second derivative of the recovered background spectrum ($T_\mathrm{b}(\upsilon)$), we would simultaneously obtain the reasonable parameters ($\tau_0$, $\upsilon_0$, and $\sigma$) \citep{Krco2008}.

From our fitting procedure, varying the $p$ value to 0.8$\leq$p$\leq$1 and the assumption of $T_{\rm k}$ from 10 to 25 K would result in an uncertainty on $\tau(\upsilon)$ of within 19\%. Varying $\tau_{\rm f}$ would not bring too many uncertainties ($\lesssim$10\% even if $\tau_{\rm f} \sim 1$), but would increase the upper limit of $T_{\rm k}$ (up to 35\,K) required to produce HINSA. Thus, we consider an overall uncertainty on $\tau(\upsilon)$ of 19\%. This is far below the uncertainty from the polynomial fitting procedures \citep[$\sim$50\%,][]{Li2003}.

\section{The 3$\sigma$ upper limit of CO}\label{sec:co}

The column density can be written as a function of optical depth \citep[$\mathrm{\tau_\nu}$,][]{Mangum2015}:
\begin{equation}
N_{\rm tot} = \frac{3h}{8{\pi}^3\left | \mu_{\rm lu} \right |^2} \frac{Q_{\rm rot}}{g_{\rm u}} \frac{e^{\frac{E_{\rm u}}{kT_{\rm ex}}}} {e^{\frac{h\nu}{kT_{\rm ex}}}-1}  \int \tau_\nu d\upsilon,
\label{eq:n_tot}
\end{equation}
where $\mathrm{\left | \mu_{lu} \right |^2}$ is the dipole matrix element, $\mathrm{Q_{rot}}$ is the rotational partition function, $\mathrm{g_u}$ is the degeneracy of upper energy level, $\mathrm{E_u}$ is the upper energy level, and $T_\mathrm{{ex}}$ is the excitation temperature.
According to the radiative transfer equation, the main beam brightness temperature $T_{\rm mb}$ is given by the expression, 
\begin{equation}
T_{\rm mb} = \eta_{\rm mb}f_{\rm b} [J(T_{\rm ex})-J(T_{\rm bg})](1-e^{-\tau_\nu}),
\label{eq:t_mb}
\end{equation}
where $\eta_{\rm mb}$ is the main beam efficiency, and $f_{\rm b}$ is the beam filling factor (assumed to be 1 in the following calculation). $T_{\rm bg}$ is the background continuum, which should be equal to the 2.73\,K for CO, and 3.9\,K for OH\,1667\,MHz including the Galactic synchrotron background. The temperature terms in equation \ref{eq:t_mb} is in the form of the Planck function: $J\mathrm{(T)}$ = $(h\nu/k)/(exp(h\nu/kT)-1)$ at millimeter wavelengths.

With the optically thin assumption ($\mathrm{\tau_\nu \ll 1}$), equation \ref{eq:n_tot} can be rewritten in the form of $T_\mathrm{b}$:
\begin{equation}
N_{\rm thin} = \frac{3h}{8{\pi}^3\left | \mu_{\rm lu} \right |^2} \frac{Q_{\rm rot}}{g_{\rm u}} \frac{e^{\frac{E_u}{kT_{\rm ex}}}} {e^{\frac{h\nu}{kT_{\rm ex}}}-1} \frac{\int T_{\rm b} d\upsilon}{J(T_{\rm ex})-J(T_{\rm bg})} .
\label{eq:n_thin}
\end{equation}

For each transition, the $\mathrm{\left | \mu_{lu} \right |^2}$, $\mathrm{E_u}$, and rest frequency $\nu$ are taken from SPLATALOGUE \footnote{www.cv.nrao.edu/php/splat/}.

We adopt a typical value measured from diffuse clouds, which is $T_\mathrm{ex}$ = 4\,K \citep{Goldsmith2013,Luo2020}.
If we fill in all the constants in equation \ref{eq:n_thin}, a convenient form of this equation is:
\begin{equation}
\begin{matrix}
N_{\rm CO} = 6.61 \times 10^{14} Q_{\rm rot}e^{\frac{5.53}{T_{\rm {ex}}}} \left [ e^{\frac{5.53}{T_{\rm {ex}}}}-1 \right ]^{-1} \\
\frac{\int T_{\rm {mb}} d\upsilon ({\rm km\,s^{-1}}) }{J(T_{\rm {ex}})-0.84} \ [{\rm cm^{-2}}].
\end{matrix}
\end{equation}

The line rms of CO within 0.5$^\circ$ regions is 5.5\,mK, and therefore the resultant 3$\sigma$ upper limit of CO column density is 9 $\times$ 10$^{13}$\,cm$^{-2}$. 

\section{PDR modeling}\label{sec:pdr}

We used the publicly available PDR code\footnote{https://uclchem.github.io/} {\sc 3d-pdr} \citep{Bisbas2012} to model the gas properties of the H\,{\sc i} arc. {\sc 3d-dpr} models one- and three-dimensional density distributions; it calculates the attenuation of the FUV radiation field in every depth point and outputs the self-consistent solutions of chemical abundances, gas and dust temperature, emissivities, and level populations by performing iterations over thermal balance (see \citealt{Bisbas2012} for more description). In the {\sc 3d-pdr} models, we used a reduced network, which contains 33 species and 330 reactions taken from the UMIST2012 network \citep{McElroy2013}. The initial gas phase abundances of He, C$^+$, and O are $9\times10^{-2}$, $1.5\times10^{-4}$ and $3\times10^{-4}$, respectively. 

\begin{figure*}
\includegraphics[width=1.0\linewidth]{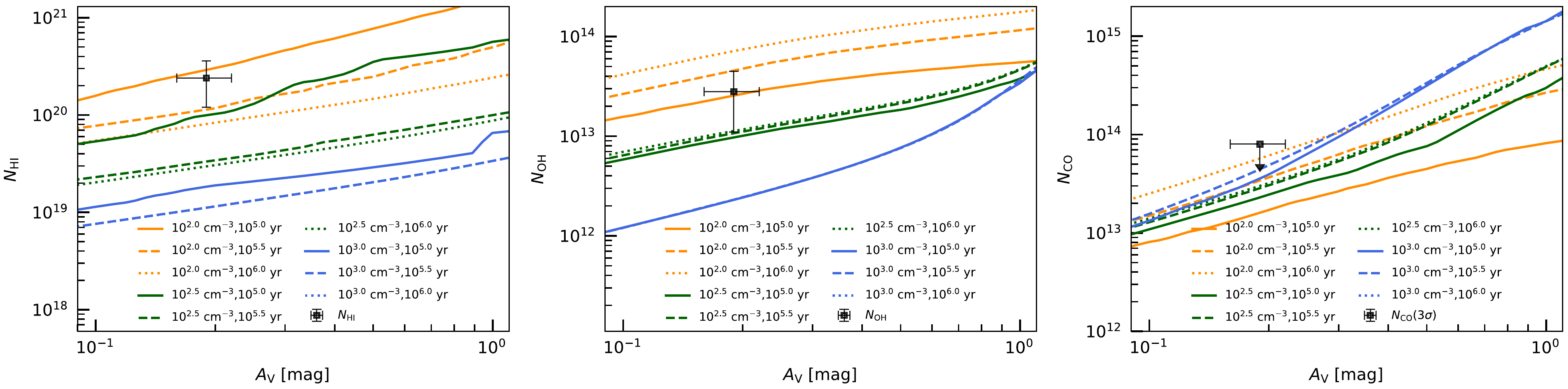}
\caption{Modeling results of $N_{\rm HI}$, $N_{\rm OH}$, and $N_{\rm CO}$ under different physical parameters (from left to right, respectively). The orange, green, and blue curves represent the evolution timescale t = 10$^5$, 10$^{5.5}$, and 10$^6$ yr. The solid, dashed, and dotted curves represent the gas densities $n_{\rm H}$ = 10$^2$, 10$^{2.5}$, and 10$^3$\,cm$^{-3}$. The black points denote data from observations.}
\label{fig:pdr}
\end{figure*}

Figure \ref{fig:pdr} shows the results under gas volume density $n_{\rm H}$ = 10$^2$, 10$^{2.5}$, and 10$^3$\,cm$^{-3}$, and evolution timescale t = 10$^5$, 10$^{5.5}$, and 10$^6$ yr. The external UV radiation field is taken as $\chi/\chi_0$ = 0.3 (normalized to the spectral shape of \citealt{Draine1978}), assuming an exponential decrease as a function of $R_{\rm GC}$ \citep{Wolfire2003}. The cosmic-ray ionization rate is taken as $1.6\times 10^{-16}$\,s$^{-1}$ (Section \ref{sec:hinsa fraction}). The metallicity is scaled with $R_{\rm GC}$ and taken to be 1/2 $Z_\odot$ \citep{Mendez-Delgado2022}. The column densities from the models are integrated along the line of sight. The observational results favor the models with a gas density in the range of $10^2 \leq n_{\rm H} \leq 10^{2.5}$\,cm$^{-3}$ and an evolution timescale in the range of $10^5 \leq t \leq 10^{6}$ yr, which is consistent with our analysis in Section \ref{sec:hinsa fraction}. This dependency on the evolution timescale was also recently addressed by \citet{Hu2021}.

The predicted CO column density is lower than the detection limit by a factor of 2$\sim$5. Given that the gas traced by CO may have a beam filling factor of $f_{\rm b} < 1$, the detection limit should be higher. On the other hand, the formation of CO in our models does not consider the enhancement due to the neutral--ion reactions through the propagation of Alfv$\acute{\rm e}$n waves \citep{Federman1996}. Such an enhancement would result in an increase in the CO column density at low $A_{\rm V}$ \citep[$A_{\rm V} < 1.5$ mag for $n_{\rm H} = 10^2$ cm$^{-3}$,][]{Bisbas2019} by up to an order of magnitude.

\end{appendix}
%
%
\end{document}